\documentstyle[aps,epsf,12pt]{revtex}
\tighten
\draft

\newcommand{\kp}{\kappa}
\newcommand{\eps}{\epsilon}
\newcommand{\beq}{\begin{equation}}
\newcommand{\eeq}{\end{equation}}
\begin{document}

\title{Orbit stability in billiards in magnetic field}
\author{Zolt\'an Kov\'acs\protect\thanks{Permanent address: 
          Institute for Theoretical Physics, 
          E\"otv\"os University, 
          Puskin u.\ 5--7, H-1088 Budapest, Hungary.}}
\address{Instituut voor Theoretische Fysica, Universiteit van Amsterdam\\
         Valckenierstraat 65, NL--1018 XE Amsterdam, The Netherlands}
\maketitle

\begin{abstract}
We study the stability properties of orbits in dispersing billiards 
in a homogeneous magnetic field by using a modified formalism based 
on the Bunimovich--Sinai curvature (horocycle method). 
We identify simple periodic orbits that can be stabilized by the
magnetic field in the four-disk model and the square-lattice Lorentz gas.
The stable orbits can play a key role in determining the transport
properties of these models.
\end{abstract}

\section{Introduction}
\label{sec-intro}

Recent experiments on ballistic transport in mesoscopic systems 
\cite{ball,antidot,antidot-2} 
have raised the interest on chaotic billiards where the addition of
an external magnetic field can yield simple models expected to account
for certain aspects of the observed behaviour. 
Some of the transport anomalies found in the experiments are thought
to have classical origins \cite{BH9,RA,Baskin,Fliesser,GKS,FGK}, 
which makes classical dynamics in these models worth studying.
In addition, the classical orbits are essential ingredients of
semiclassical calculations intended to account for these effects.
A powerful formalism can be based on periodic orbits and their stability
properties \cite{cycle-exp,semi-cyc,vgdet}; 
moreover, classical quantities characterizing the billiard and
apperaring in various formulae describing transport properties depend
very strongly on the hyperbolic or nonhyperbolic nature of the billiard.
These considerations indicate that calculating the stability of orbits
and locating stable islands in phase space for these models can be of
central importance.

The geometry of experiments done so far on microjunctions 
\cite{ball} and antidot-lattices \cite{antidot,antidot-2} can be
imitated by models consisting of circular scatterers (disks). 
The area between four disks centered on the corners of a square
(four-disk model) has an adequate geometry for a cross-junction while
the square-lattice Lorentz gas \cite{lg0} corresponds to an
antidot lattice with a periodic arrangement of scatterers.
In fact, the four-disk model has been studied in connection with
cross-juction experiments  \cite{BH9,RA},
and transport in the Lorentz gas in magnetic field have been 
investigated to explain some of the magnetoresistance phenomena 
in antidot-lattices  \cite{Baskin,Fliesser}.
For the above reasons, we will focus on these hard-disk billiards. 

Concerning the role of stable orbits, the three-disk model in magnetic
field has already been studied from the point of view of chaotic
scattering, demonstrating the effect of stable islands created by the
field on the global properties of the dynamics   \cite{3disk-magn}.
Closed billiards like the stadium in an external magnetic field have
been investigated in Ref.~\cite{tg-map} where the form of the 
tangent map for billiards in magnetic field has also been given.
In the present paper, we propose an alternative tool, namely the
extension of the horocycle method of Bunimovich and Sinai
\cite{kappa0} worked out originally for the calculation of Lyapunov
exponents in Euclidean billiards (it has also been applied to
billiards on a surface of constant negative curvature \cite{cnc-bill}).
This modified method is considerably simpler to implement than
calculations based on tangent maps and makes it easy for us to find
stabilized orbits and follow their stability properties as a function
of the field strength.

\section{Disk models in magnetic field}
\label{sec-models}

We consider the motion of a charged particle of unit mass in a plane
colliding with hard disks of radius $r$ in the presence of a uniform
external magnetic field $h$ perpendicular to the plane;
the velocity of the particle is fixed at $v=1$.
In addition we set the unit of electric charge so that the cyclotron
radius is simply $s=1/h$ and choose the signs of the charge and the
field so that the particle goes around counterclockwise along the 
cyclotron orbit.
The disk centers are in the vertices of a square lattice with a period
$2d=2$; the whole lattice filled with disks yields the (periodic)
Lorentz gas, while taking a single square only we get the four-disk model.
The geometry of the system is controlled by the radii $r$ and $s$.
Throughout the paper, we will use the cyclotron radius $s$
measuring the strength of the magnetic field as our control parameter
while we keep the disk radius constant at $r=1/2$. 
All the numeric values of $s$ will refer to this case and
we will comment on the dependence of our findings on $r$ when necessary.

During the free motion the particle follows a cyclotron orbit of
radius $s$ around a given center depending on the initial conditions;
a collision (considered specular) with a disk puts the particle onto
another circular path with the same radius but around a different center.
The dynamics as a sequence of collisions can be kept track of by
following the center positions of these circles \cite{3disk-magn}.
A jump in the center position always occurs when the particle collides
with a disk.
Thus, the sequence of the circle centers defines a discrete mapping
(a Poincar\'e map) associated with the time-continuous dynamics. 
Let $(p,q)$ denote the orbit center at a given instant of time; after
collision, a new center $(p',q')$ defines the particle trajectory.
The Poincar\'e map $(p',q') = P(p,q)$ connecting these coordinates 
relative to the position of the disk on which the collision occurs 
has been given explicitely in Ref.~\cite{3disk-magn}; we do not need 
its precise form here.

The complete dynamics can be described by subsequent 
applications of the map $P$ with the proper disk center coordinates.  
$p$ and $q$ are canonically conjugate pairs, and the plane 
$(p,q)$ can be taken as the phase space of our system. 
They can also be used in conjunction with the disk centers to compute
other quantities useful in our study.
For example, we will need the impact angles at collisions and the
lengths of the arcs between them; they are easy to obtain from the
center coordinates.
The tangent map of the process can be calculated as the derivative of
$P$ with respect to $p$ and $q$ \cite{3disk-magn}, and the result is
equivalent to the map given in Ref.~\cite{tg-map}.

\section{Lyapunov exponents and horocycles}
\label{sec-kappa}

The stability of a periodic orbit in the Poincar\'e section
is described by its two Lyapunov
exponents; due to the Hamiltonian character of the billiard models,
the two exponents add up to zero. 
Hyperbolic orbits have real Lyapunov exponents so the largest one is
positive describing the exponential divergence of nearby
trajectories, while elliptic orbits are characterized by
imaginary stability exponents connected to their winding numbers.
The Lyapunov exponent of the cycle can be calculated as the logarithm
of the eigenvalue of its stability matrix obtained from the tangent map.
The eigenvalues are negative when there is an odd number of inversion 
of nearby orbits during the cycle; then we take the modulus first.
We are mainly interested in the {\em evolution} of stability
properties of periodic orbits when changing the control parameter, 
but the fact that the stability matrix is usually a product of tangent map
matrices, each describing one collision, makes it rather difficult  
to obtain considerable insight to orbit stability from the
complicated form of the stability matrix elements.
To avoid these difficulties, we will apply in this paper 
a more suitable approach, which is
based on the curvature of local orbit fronts.

\subsection{The curvature $\protect\bbox{\kappa}$}

Making use of the local divergence property, there is another possibility
for the calculation of Lyapunov exponents.
Its central quantity is the Bunimovich--Sinai curvature $\kp$, 
which is the local curvature of the front (horocycle) formed around a
central trajectory by nearby orbits started from a common point of
origin \cite{kappa0}.
Its evolution in time $\kp(t)$ along the central orbit can be given
easily: it is a smooth function with jumps due to the collisions. 
Without an external field, the local radius of the front grows
linearly in time, so between collisions the curvature simply evolves as
\beq
          \kp (t) = \frac{\kp_0}{1 + \kp_0 t}    
\label{eq-kp0-free}
\eeq
with $\kp_0$ being the curvature at $t=0$. In a collision, $\kp$
jumps from its value $\kp_-$ immediately before hitting the disk to a
new value $\kp_+$ connected by
\beq
          \kp_+ = \kp_- + \frac{2}{r \cos \phi}  \  ,
\label{eq-kp0-jump}
\eeq
where $r$ is the disk radius and $ - \pi /2 < \phi < \pi /2 $ is the 
impact angle formed by the direction of the incoming velocity and the
inward normal of the disk at the point of impact.
Since this jump is always positive, starting from $\kp_0 > 0$
(diverging orbits), the curvature remains positive for all time.

The Lyapunov exponent per collision of an orbit is given as
\beq     
     \lambda = 
        \lim\limits_{N \rightarrow \infty} 
              \frac{1}{N} \int\limits_0^{\: T_N}  \kp(t) \, dt \  ,
\label{eq-lyap}
\eeq
where $T_N$ is the time at the $N$th collision along the orbit.
In the long-time limit, the dependence on the initial curvature $\kp(0)$
is expected to disappear in the same way as the dependence on initial
conditions in any other method for calculating Lyapunov exponents.

For a periodic orbit with $n$ collisions, the integral can be taken
over one time period $T$ with a periodicity condition
implied on $\kp(t)$ so that $\kp(T) = \kp(0)$.
It is easy to show that there are two solutions, one positive and the
other negative, for this requirement.
If we denote the values of the front curvature right after the $m$th 
collision and the next one by $\kp_m$ and $\kp_{m+1}$, respectively,
then it follows from Eqs.~(\ref{eq-kp0-free}) and (\ref{eq-kp0-jump}) 
that they are connected by a rational function with positive coefficients.
This means that in the periodicity condition $\kp_n = R(\kp_0) = \kp_0$ 
the rational function $R$ also has this property and thus we obtain
two solutions with opposite signs.
Obviously, the positive value of $\kp_0$ leads to the positive
exponent when evaluating the integral.
The corresponding eigenvalues of the stability matrix are
$\Lambda_{\pm} = (-1)^n \exp (\pm\lambda n)$, where the sign factor in
front of the exponential accounts for the inversion of nearby orbits
occuring at each collision.

It is important to note that the curvature of the front can also be
written as a ratio by using the local Jacobi coordinate $J$, which is
the (signed)
distance of a nearby orbit from the central one along the front
\cite{Jacobi-Ricatti}.
The inverse of $J$ multiplied by the angle between the two
velocities gives just $\kp$. 
Since we set the velocity $v=1$, this (small) angle is equal to the
velocity component of the nearby orbit in the direction perpendicular to that
of the central one, thus it can be written as the time derivative $\dot{J}$.
Therefore $\kp = \dot{J}/J$, and since $\dot{J}$ and $J$ can be considered
as components of a vector in the tangent space of the dynamics, 
this means that $\kp$ is determined by the direction of that vector.
This relationship also implies that 
the free-flight evolution of $\kp(t)$  satisfies the Ricatti equation:
\beq 
      \dot{\kp} = - K - \kp^2 \  ,
\label{eq-Ricatti}
\eeq
where the curvature term $K$ follows from the Jacobi equation 
describing the separation of geodesics: 
\beq 
       \ddot{J} + K J = 0 \  .
\label{eq-Jacobi}
\eeq
In our case, $J(t)$ is a linear function, which corresponds to 
$K=0$ in Eqs.~(\ref{eq-Ricatti}) and (\ref{eq-Jacobi}) as expected for
a Euclidean billiard.
We also note that the situation of crossing orbits means $J$ changing sign
while going through 0, which leads to a $1/t$ type singularity in $\kp(t)$.
Such an inversion is accompanied by a sign change in the stability
eigenvalue, so we may complete the sign rule for the eigenvalues: each
jump (at collision) and each pole singularity in $\kp(t)$ contributes a 
factor $-1$ to the eigenvalues.

\subsection{$\protect\bbox{\kappa}$ in magnetic field}

The presence of the external field modifies the evolution of $\kp$
both in the free motion and at the collisions. 
The fact that $\kp(t)$ can be expressed through the Jacobi coordinate
$J(t)$ makes it easy to calculate it in a magnetic field.
During the free flight, the particle moves along a circle of radius $s$.
Starting two orbits at $t=0$ with parallel velocities, their distance
remains constant along the direction perpendicular to the 
{\em initial} velocity while the tangent of the front turns with the 
velocity direction along the orbits.
The local coordinate $J(t)$ is the projection of the 
constant distance $J_0$ on the front direction at time $t$: 
$J(t) = J_0 \cos (t/s)$.
By moving the origin of time to the point where the orbits cross, the
cosine is replaced by sine; with this choice the free evolution of the
curvature can be written as
\beq
          \kp (t) =  \frac{1}{s} \: \cot \frac{t}{s} \   .
\label{eq-kp1-free}
\eeq
This form has exactly the same singularity ($1/t$) approaching the
points of orbit crossing as in the fieldless case.
It is also periodic with a period equal to half of the total turnover 
time $T_c = 2 \pi s$ along the cyclotron orbit.
It is important to stress that $\kp(t)$ can have {\em negative} values
as well corresponding to the focusing effect of the magnetic field.

The jump in $\kp$ at collisions comes from the fact that there is a 
sudden change in the relative direction of two nearby orbits due to
their different impact angles.
In the fieldless case, the change in the relative direction 
is just twice the angle difference between the normal directions at 
the points of impact on the disk perimeter.
With magnetic field, there is an additional effect due to the constant
change in the velocity direction of the orbits during free flight: 
If the central orbit of the front reaches the disk under an impact
angle $\phi$, then another orbit in the front with a local Jacobi 
coordinate $J$ will have a relative velocity direction different from 
$\dot{J}$ when colliding with the disk because of the time difference 
$\delta = J \tan \phi$ between the two impacts.
This leads to an additional jump $2 \delta /s$ in $\kp$ at the collision.
As a result, we obtain a modified jump formula:
\beq
  \kp_+ = \kp_- + \frac{2}{\cos\phi} 
                \left( \frac{1}{r} + \frac{1}{s} \sin\phi \right) \ .
\label{eq-kp1-jump}
\eeq
The parenthesized part can be interpreted as an effective curvature of
the scatterer for the orbits: the field changes it from its fieldless
value $1/r$ so that even a collision with a flat wall causes a jump in
$\kp$ in magnetic field. 
The field can decrease (increase) the defocusing effect of a collision
on the disc at negative (positive) impact angles; the collision can
even be focusing if $s < r$.
It is worth emphasizing that both the modified free evolution and the
jump condition simplify, as expected, to the fieldless formulae 
(\ref{eq-kp0-free}) and (\ref{eq-kp0-jump}) in the limit 
$s \rightarrow \infty$.

In the following, we will use the normalized time variable
$\tau =t/s$ and its function $\xi(\tau) = s \kp (\tau s)$, which we
call the {\em curvature function\/} of an orbit.
It consists of pieces of the cotangent function 
$\cot (\tau - \tau_i + \eps_i)$ $(i=0,1,2,\ldots)$ 
describing the arcs of free flights and positioned by
the phase shifts $0 \le \eps_i < \pi$ 
so that the jumps due to the collisions at
$\tau_i$ satisfy the condition (\ref{eq-kp1-jump}) prescribed for $\kp$
(now obviously multiplied by a factor $s$ for $\xi$).
We will also use the notation $\xi_i=\xi(\tau_i +0) = \cot \eps_i$.
For a periodic orbit of $n$ collisions, we will assume that the 
periodicity condition $\xi_0=\xi_n$ is imposed when possible; 
in addition, we choose the origin of time so that the phase shift of
the first piece is $\eps_0= \tau_0$, 
i.e., we move the initial collision when we start following the orbit
to the value $\tau_0 = \cot^{-1} \xi_0$.
This way we can compare the curvature functions of periodic orbits at
different values of the magnetic field.

\subsection{Lyapunov exponents of periodic orbits}

The Lyapunov exponent $\lambda$ of a cycle of period $T$ with
$n$ collisions can now be obtained as
\beq
      n \lambda = \int\limits_{\displaystyle{\tau_0}}^{\displaystyle{\tau_n}} 
                        \xi(\tau) \, d\tau \  ,
\eeq
where $\tau_n = \tau_0 + T/s$. 
The integral can easily be evaluated over the continuous pieces, but
we can also combine the calculation of $\lambda$ with the solution of 
the periodicity condition. 
For this, it is enough to notice that by using elementary
trigonometrics we can still write the relationship between 
$\xi_i$ and $\xi_{i-1}$ as a rational function:
\begin{eqnarray}
   \xi_i & = & \cot (\eps_{i-1} + \alpha_i) + \Delta_i \nonumber \\
         & = & \frac{
     (\cot \alpha_i + \Delta_i) \xi_{i-1} + (\Delta_i \cot \alpha_i -1) }
         { \xi_{i-1} + \cot \alpha_i }  \nonumber  \\[.5ex] 
         & = & \frac{
     (\cos \alpha_i + \Delta_i \sin \alpha_i) \xi_{i-1} 
                             + (\Delta_i \cos \alpha_i - \sin \alpha_i) } 
         { \xi_{i-1} \sin \alpha_i + \cos \alpha_i } \  ,
\label{eq-xi-rat}
\end{eqnarray}
where $\alpha_i=\tau_i - \tau_{i-1}$ is the length of arc $i$ measured 
in angle and $\Delta_i$ is the jump in $\xi$ at $\tau_i$.
We wrote the third equality using normalized coefficients 
with a determinant 1. 
This means that the periodicity condition can also be written 
in the form
\beq 
      \xi_0 = \frac{A\xi_0 + B}{C\xi_0 + D}
\label{eq-xi0}
\eeq
with a determinant $AD-BC=1$.
This is a quadratic equation for $\xi_0$, and
it is easy to check that the two eigenvalues of the matrix 
$ M = \left( \begin{array}{rr} A  &  B \\
                               C  &  D  \end{array}  \right) $
are just equal to the denominator in Eq.~(\ref{eq-xi0}) 
taken with the two solutions for $\xi_0$.
Using the connection between $\kp$ and the Jacobi coordinate $J$, 
it can be shown that the matrix $M$ is equivalent to the stability matrix
of the cycle
obtained from the tangent map, so the Lyapunov exponents are simply
$n \lambda = \log |C \xi_0 + D|$.

We can immediately see that the periodicity condition can be satisfied
if and only if the orbit is hyperbolic, since the two solutions of
$\xi_0$ yield the two Lyapunov exponents.
However, now the coefficients are not necessarily positive, thus in
magnetic field we have the possibility of {\em elliptic} cycles with 
complex eigenvalues when Eq.~(\ref{eq-xi0}) has no real solution.
For these stable orbits, $\xi(\tau)$ is not periodic and the connection
between its integral and the complex eigenvalues becomes more complicated.
It could be restored by a formal extension allowing complex values for
$\kp$ and $\xi$, but there is a more plausible interpretation that we
will discuss later.

\section{Orbit stabilization in magnetic field}
\label{sec-orbits}

The positivity of $\kp(t)$ in the fieldless case ensures that all
the periodic orbits in the hard-disk billiards are hyperbolic. 
Small field values ($s \gg 1$) are not expected to change that: 
the angles $\alpha_i$ of the arcs are small, so the coefficients in 
Eq.~(\ref{eq-xi0}) remain positive.
Turning to a stronger field, there can be periodic orbits with wider
negative parts in their curvature function leading 
to a decreased instability. 
The cycle becomes marginally stable at the field value for which the
integral of $\xi(\tau)$ vanishes. 
Increasing the field further, we reach the region of elliptic stability 
with imaginary Lyapunov exponents, and at another critical field strength 
the cycle disappears through an inverse saddle-center bifurcation. 
This process, already observed in Ref.~\cite{3disk-magn}, 
can be studied best on the simplest periodic orbits of the system.

\subsection{The clockwise orbit}

Based on the form of Eq.~(\ref{eq-kp1-free}), one expects that the
candidates for stabilization are the orbits consisting in strong fields
mostly of arcs close to a semicircle so that the corresponding pieces
in the curvature function stretch almost over the whole period of the
cotangent function.
The simplest case is a cycle containing arcs of equal length and
identical collisions ensured by the symmetry of the system. 
We take the four-disk model as our first example. 
Each disk has a symmetry point, the section of the disk perimeter with 
the diagonal of the square.
There are two period-4 orbits connecting all these points
one after the other in a clockwise or counterclockwise direction. 
Without magnetic field, these two orbits coincide on the configuration
plane, forming a square; taking the fourfold symmetry into account, they are 
essentially fixed points of the symmetry-reduced version of the map 
$P$ acting on the arc centers.

\begin{figure}
\epsfxsize=8cm
\centering\leavevmode\epsfbox{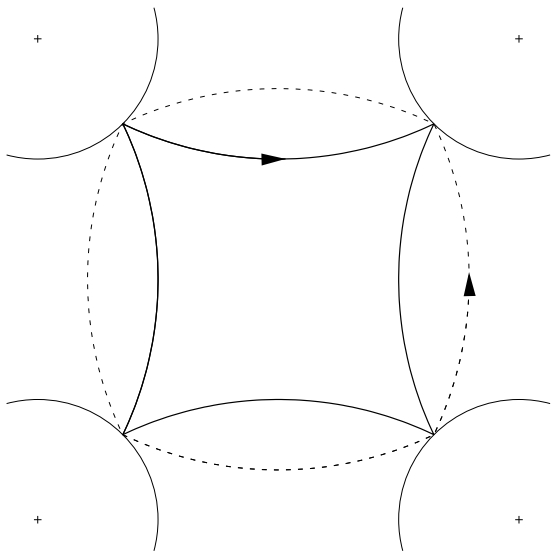}
\caption{The clockwise (heavy line) and counterclockwise (dashed line)
orbits in the four-disk model for $r=0.5$ and $s=1.5$; 
the crosses mark the disk centers.}
\label{fig-orbcw}
\end{figure}
\begin{figure}
\centering\leavevmode \epsfysize=6cm\epsfbox{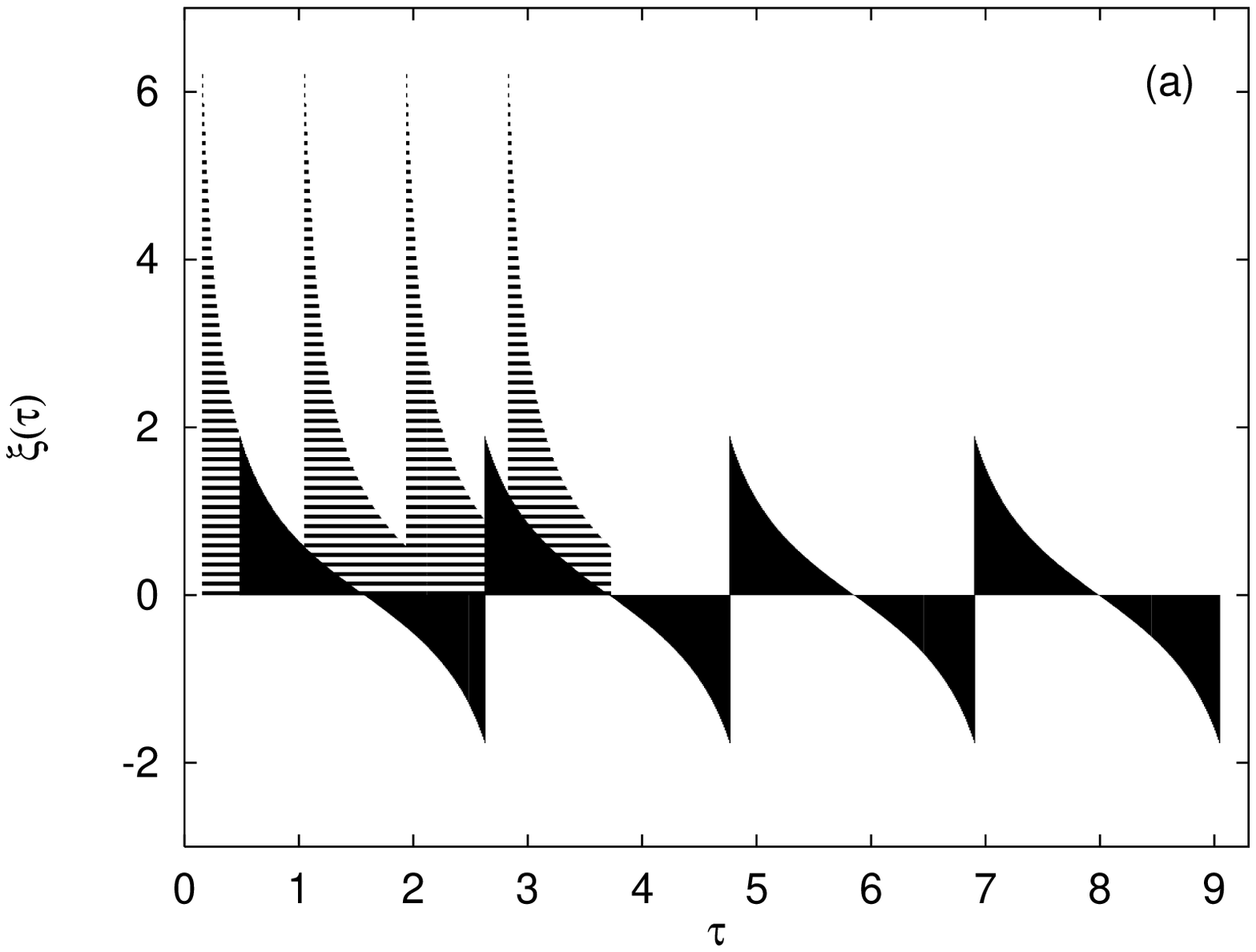} \\[.5cm]
\centering\leavevmode \epsfysize=6cm\epsfbox{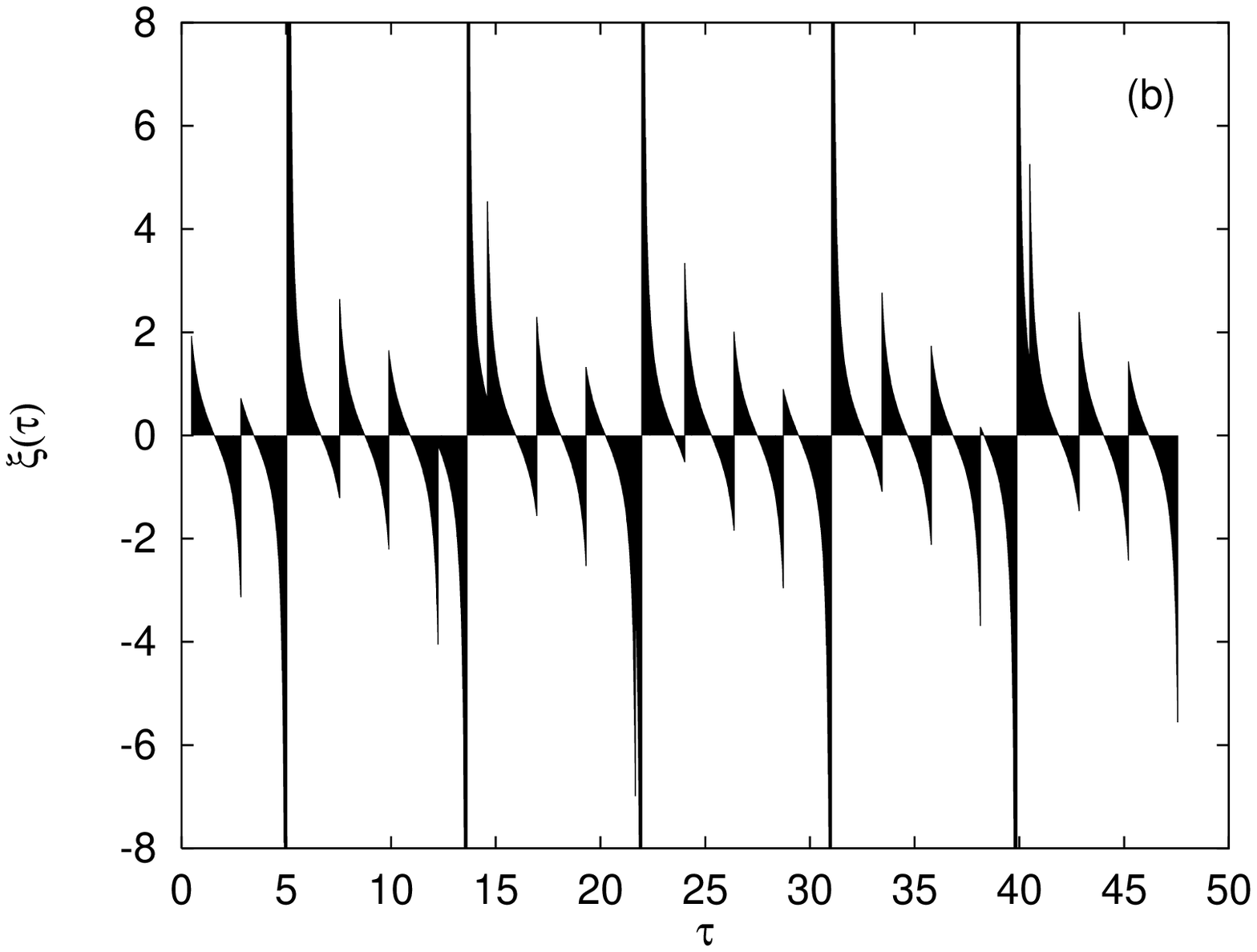} \\[.5cm]
\centering\leavevmode \epsfysize=6cm\epsfbox{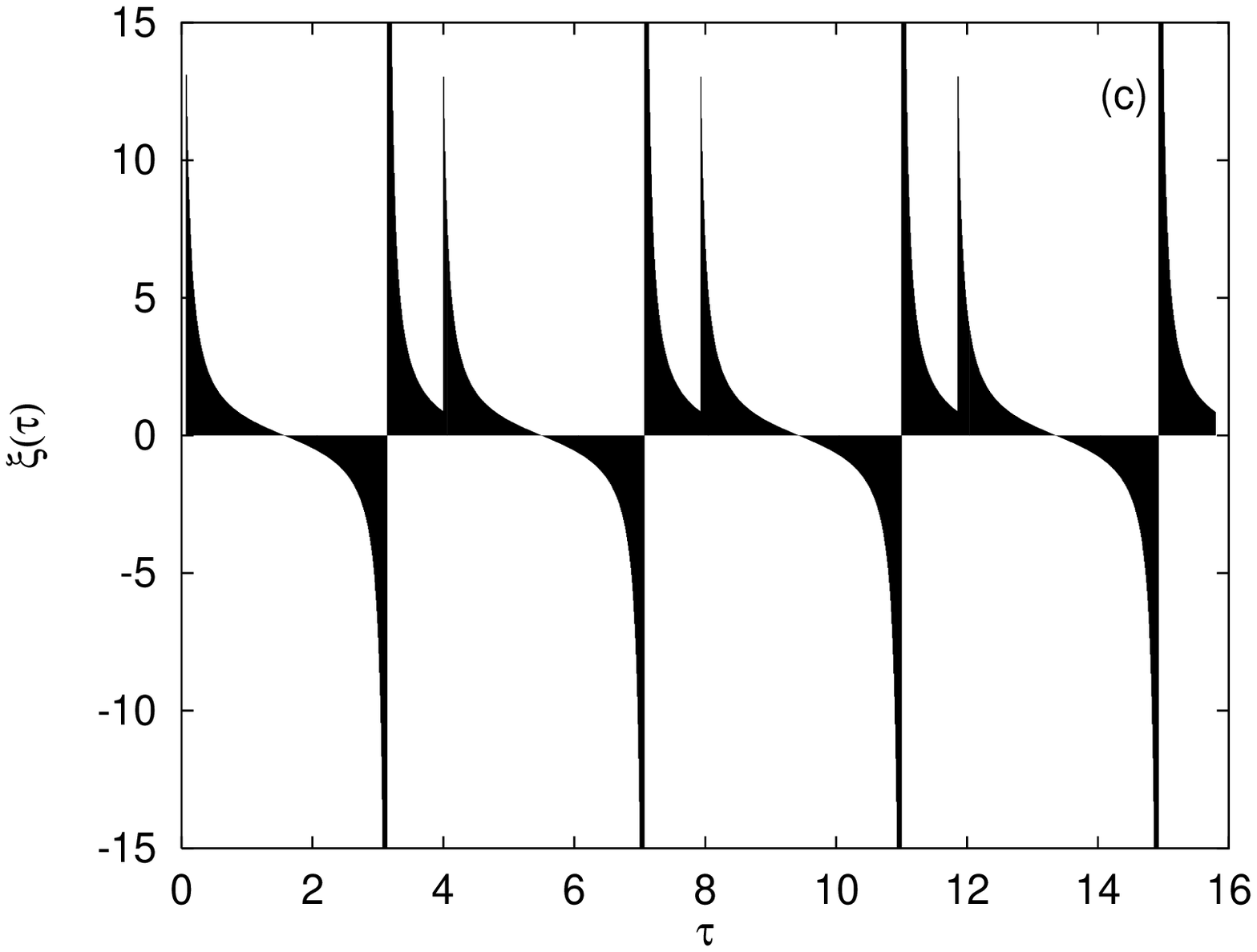}
\caption{The curvature function $\xi(\tau)$, plotted with vertical lines 
drawn from the axis $\xi=0$, of the clockwise orbit in the hyperbolic 
regime (a) at cyclotron radii $s=.737 \approx s_1$ (full lines) and 1.5 
(dashed lines) for a full period of four collisions; 
$\xi(\tau)$ of the clockwise orbit in the elliptic regime (b) at $s=0.7$
for 5 full periods (20 collisions); the curvature function of the 
{\em saddle} clockwise orbit (c) at $s=0.7$ for a full period.
Notice the difference of scales in the plots.
The curves of the first blocks of the two cases in Fig.~2a coincide because
of the normalization of $\xi(\tau)$.
In Fig.~2b, there is approximately one pole singularity for each
period of four collisions due to the fact that the phase of the complex
eigenvalue is close to $3\pi/4$ (cf.\ Eq.~(\protect\ref{eq-phi})).}
\label{fig-kpcw}
\end{figure}

Turning on the field, the straight lines in the cycle are replaced by 
arcs and the orbit pair splits up (Fig.~\ref{fig-orbcw}). 
It can be checked, e.g.\ by direct calculation using the tangent 
map as in Ref.~\cite{3disk-magn}, that the largest Lyapunov exponent of the 
clockwise (counterclockwise) orbit decreases (increases) with
increasing field strength. 
The fourfold symmetry remains, and the curvature function of the full
orbit consists of four copies of the block describing free flight
along an arc and the jump at the end of it (Fig.~\ref{fig-kpcw}a); 
in other words, the periodicity condition can just be applied to this
block due to the symmetry as if we had a period-1 cycle.

In the clockwise orbit, the jump at the collision decreases 
with increasing field strength, while the length of the arc increases 
leading to negative regions appearing in $\xi(\tau)$.
For a cyclotron radius $s_1$, we arrive at a situation
(Fig.~\ref{fig-kpcw}a) where the jump
$\Delta$ is just equal to the {\em minimum} possible difference
between two values of the cotangent function at points separated by a
distance $\alpha$, the angular length of the arcs:
\beq
    \Delta = 2 \tan \frac{\alpha}{2} \ .
\label{eq-s1-clock1}
\eeq
Because of the symmetry of the cotangent function, the integral of
$\xi(\tau)$ over the block is zero, i.e., the orbit is marginally stable.
By writing $\Delta$ and $\alpha$ as functions of $s$,
Eq.~(\ref{eq-s1-clock1}) can be used to calculate the critical field
value:
\beq
     s_1 = \sqrt{ \left( 1 - r/\sqrt{2} \right) ^2 
                + \left( r/\sqrt{2} \right) ^2 } 
\eeq
(remember that $d=1$).
This means that the clockwise cycle reaches marginal stability, 
at $s_1 \approx 0.7368$ for $r=0.5$, when the center of an arc in the cycle 
falls in line with the centers of the disks connected by the arc.

\subsection{The elliptic regime}

For field values $s < s_1$, we are in the elliptic regime and 
the periodicity condition (\ref{eq-xi0}) cannot be satisfied 
by any real value of $\xi_0$.
The connection to the complex eigenvalues $e^{\pm i\varphi}$ of the
stability matrix could then be restored by a formal extension allowing
complex values for $\kp$ and $\xi$, but there is also a more plausible
interpretation based on the dynamics unfolding from the sequence of
phase shifts $\eps_i$ in the blocks of the curvature function.
To see this in details, let us follow the clockwise orbit over a
large number of repetitions and record the sequence
$\eps_i=\cot^{-1} (\xi_i)$ ($i=0,1,2,\ldots$) starting
from a certain $\eps_0$ value.
It follows from the interpretation of $\kp$ as the ratio of the two
components of a vector in tangent space that the values $\eps_i$
are just the angles (taken with mod $\pi$) of these vectors in
the Poincar\'e section defined by the collisions and parametrized
by $\dot{J}$ and $J$.

The connection between $\eps_i$ and $\eps_{i-1}$ can easily be obtained
from the first line of Eq.~(\ref{eq-xi-rat}).
Since now each block has the same width $\alpha$ and jump $\Delta$
at its end, this relationship does not depend on $i$ and 
can be written as a one dimensional map: 
\beq
      \eps_i = C(\eps_{i-1}) 
               = \cot^{-1} \left( \cot (\eps_{i-1} + \alpha) + \Delta \right)
\eeq
It has the topology of a circle map on the interval
$[0,\pi)$, since the cotangent function is periodic by $\pi$.
The sequence $\eps_i$ then follows from the dynamics of this
circle map.
Recalling that the members of this sequence are angles of vectors 
connected by the stability matrix of the cycle in the Poincar\'e section,
we conclude that the circle map $C$ represents the effect of the
Poincar\'e map $M$ restricted to the directions of vectors. 

The map $C$ is essentially a shift by $\alpha$ distorted by the 
presence of $\Delta$. 
In the hyperbolic regime, where $ \Delta < 2 \tan (\alpha /2)$,
it has two fixed points, one stable and the other unstable, connected
to the two solutions of the periodicity condition for $\kp$.
For an elliptic cycle, however, it has no fixed points, so $C$ generates a
quasiperiodic sequence of the $\eps_i$ values, and one can also 
easily check that for $0 < \eps < \pi - \alpha$ the difference 
$C(\eps) - \eps$ is bounded from below. 
This implies that for any starting $\eps_0$ there must be a 
certain step $j$ so that $\eps_j + \alpha > \pi$,
i.e., the corresponding block in the curvature function will contain the 
pole singularity of the cotangent function (Fig.~\ref{fig-kpcw}b).

If we look for the total angle $\Theta_n$ by which a vector is rotated under 
$n$ applications of the Poincar\'e map $M$, then we must add to
the difference $\eps_n -\eps_0$ given by the map $C$ an extra
$\pi$  ``lost'' in the circle map for each $\eps_i > \pi - \alpha$,
i.e., for each pole singularity in $\xi(\tau)$:
\beq
    \Theta_n = \eps_n - \eps_0 + k_n \pi    \ , 
\eeq
where $k_n$ is the number of poles in $n$ blocks.
Formally, the extra term can be considered as the contribution 
of the poles to the {\em imaginary} part of the integral of $\xi(\tau)$
taken over $n$ blocks.
The real part of the integral, which can be evaluated by excluding
small boxes centered on the poles, will fluctuate in $n$ 
as a consequence of the quasiperiodic dynamics in the $\eps_i$ values, 
but for large $n$ the elliptic property of the cycle ensures that 
it vanishes when divided by $n$.
Thus we may conclude that the integral of the curvature function 
yields the Lyapunov exponent of a periodic orbit also in the elliptic regime 
in the sense that if we take it over a large number of repetitions of the 
orbit, then the phase of the eigenvalues $e^{\pm i \phi}$ is determined by
the average number of poles per collision in the orbit:
\beq
   \phi = \lim\limits_{n \rightarrow \infty} 
                             \left( 1 - \frac{k_n}{n} \right) \pi \ ,
\label{eq-phi}
\eeq
where we took into account the effect of the jump in $\kp$ at the 
collision.
The argument above and its conclusion could obviously be generalized
to cycles of longer periods too.

\subsection{Bifurcations and pruning}

At another critical field value $s_0 = 1 - r/\sqrt{2} \approx 0.6464$,
the arcs become complete semicircles, and for $s < s_0$ the cycle 
cannot exist any more.
The curvature function for $s_0$ is just the perfect cotangent
function for all times: $\xi (\tau) = \cot \tau$.
The jumps induced by the collisions sit right on top of the poles
at $\tau_n = n \pi$; thus in the integral of $\xi(\tau)$
there will be an additional sign factor for each block.
This means that approaching $s_0$ from below (decreasing field
strength) the clockwise orbit is born at $s=s_0$ with the stability  
eigenvalue $\Lambda =1$: the singularity in each block of $\xi(\tau)$ 
compensates for the sign change due to the collision.

The birth of the elliptic cycle is a usual saddle-center bifurcation
as in smooth Hamiltonian systems: it is born together with a
hyperbolic cycle.
This orbit has the same symmetry but the length of its arcs is 
{\em increasing} for $s > s_0$, so they become longer than a semicircle.
The corresponding block in the curvature function contains more 
than a full period of the cotangent function, so the periodicity 
condition can be met and the cycle is hyperbolic (Fig.~\ref{fig-kpcw}c).
By increasing $s$ further, the saddle orbit disappears in a way 
typical in hard-wall billiards: it reaches the stage where the arcs
should cut through the disks to arrive at the desired symmetry points.
This illustrates the main mechanism for {\em pruning} \cite{pruning} 
(disappearance of orbits by varying a system parameter) 
in our billiard models: the orbit becomes forbidden when an
arc hits a disk that is not in its ``itinerary'' or when a collision
angle becomes flat \cite{rem-coin}.

In the elliptic regime, the clockwise orbit in phase 
space is in the middle of a stable island 
formed by periodic, quasiperiodic, and 
chaotic trajectories as in general Hamiltonian systems.
Outside the outermost KAM torus surrounding the island, there is the 
well-known infinite hierarchy of chains of smaller islands around longer
periodic orbits, also stabilized by the field, and cantori, the remnants
of broken KAM tori; 
we will discuss these aspects of the stabilization later in more details.
It should be emphasized that the main steps of the above scenario for
the stabilization of the clockwise orbit do not depend on the disk
size relative to the disk spacing: the saddle-center bifurcation takes
place at any value of the ratio $r/d$, therefore the orbit
stabilization occurs even at small disk sizes where the fieldless
orbits are very unstable.
The disk size affects, however, the range of stability:  for small
disk radii, it is proportional to $r$ (the geometric interpretation 
given earlier for $s_1$ illustrates this point very clearly). 

In the four-disk model, the clockwise orbit is the only important example
of the stabilization induced by the field.
The two other candidates with a simple symmetry (orbits bouncing
forth and back between two disks along the side or the diagonal of 
the square connecting the disk centers) have eigenvalues independent
of the magnetic field: the focusing effect of the field along the arcs
is exactly cancelled by the additional defocusing effect at the collisions.

\section{Stable orbits in the Lorentz gas}
\label{sec-stlg}

By extending the four-disk model periodically on the plane, we obtain
the (square-lattice) Lorentz gas for which the four-disk model serves
as a unit cell.
This means, obviously, that the stable clockwise orbit of the
four-disk model is present in the Lorentz gas too. 
However, the extension creates various other orbits that can be
stabilized by the field: there are localized orbits extending over
more than one unit cell as well as nonlocalized (travelling) orbits
proceeding at a constant average speed.

\subsection{Symmetric cycles}

\begin{figure}
\centering\leavevmode\epsfxsize=14cm\epsfbox{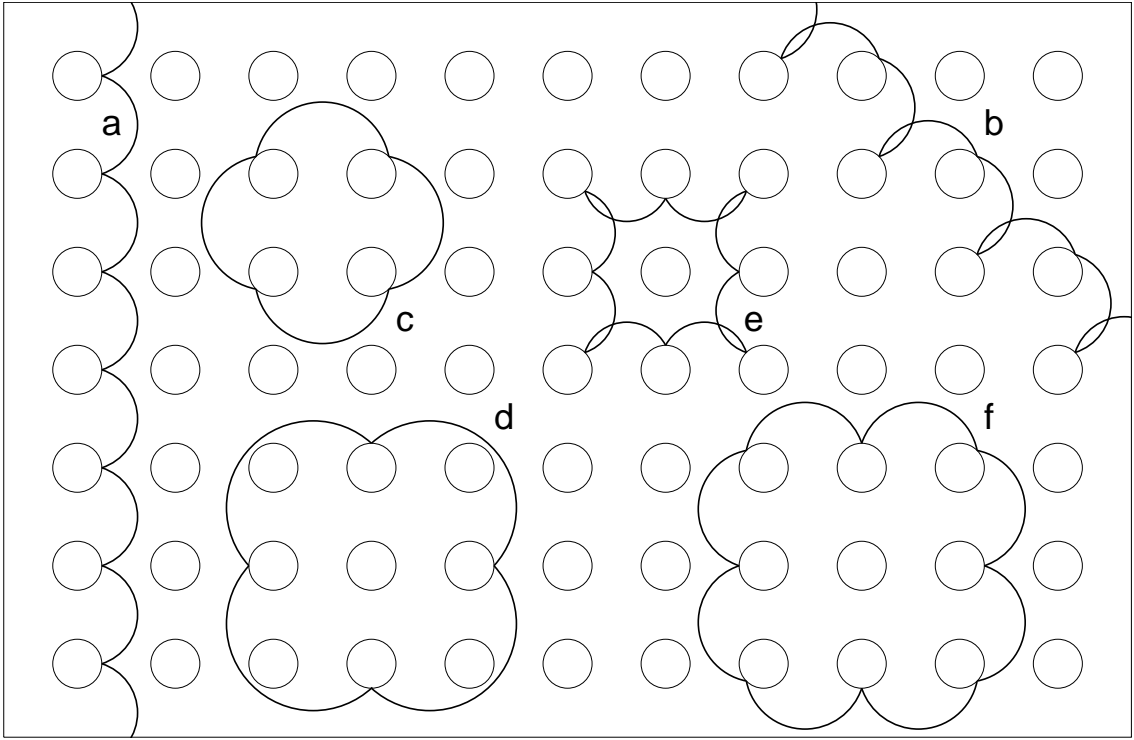}
\caption{Symmetric stable cycles in the Lorentz gas; the cyclotron
radii are $s=1.05$ (a and b), 1.38 (c), 1.77 (d), 0.9 (e), and 1.22 (f).}
\label{fig-orblg1}
\end{figure}

As we have seen, the focusing property of the magnetic field may lead
to stabilization if the arcs building up the orbit are close to
semicircles, so first we will look for orbits with this property.
The symmetries of the system can help us again to find the simplest 
stable cycles: those with arcs of equal length.
Each disk now has eight symmetry points due to the four symmetry axes,
and connecting two such points by an arc generates a symmetric cycle 
with period 1, 2, 4, or 8 (period-1 and period-2 cycles if we take the
fourfold symmetry into account). 
The configurations not prohibited by the pruning rules mentioned
earlier are all born stable 
and have their ranges of elliptic stability.
The most important examples are shown in Fig.~\ref{fig-orblg1}, 
while Table I lists the corresponding field values $s_0$ and $s_1$.

%
\begin{table}
\caption{The critical field values $s_0$ and $s_1$ of the symmetric
cycles in Figs.~1, 3, and 5 (the labels in the first column refer to
the figures) for disk size $r=0.5$. The numbers are rounded to four
decimal digits.
The $s_1$ value for the cycle 3d is in parentheses 
since it cannot be reached because of the pruning.}
\begin{tabular}{ldd}
   Cycle &   $s_0$    &   $s_1$  \\
\hline
   1     &  0.6464    &  0.7368  \\
   3a    &  1.0000    &  1.1180  \\
   3b    &  1.0000    &  1.0607  \\
   3c    &  1.3536    &  1.3990  \\
   3d    &  1.7678    & (1.8028) \\
   3e    &  0.8265    &  0.9309  \\
   3f    &  1.1791    &  1.2543  \\
   5a    &  0.6105    &  0.6197  \\
   5b    &  0.7792    &  0.8363  \\
   5c    &  0.5586    &  0.5711 
\end{tabular}
\label{T1}
\end{table}
The simplest of these cycles is the travelling period-1 orbit that
connects all the disks along a lattice axis (Fig.~\ref{fig-orblg1}a).  
It is not affected by pruning except for large disk sizes ($r > 2/3$).
It is born at $s_0 = 1$ and becomes hyperbolic at 
$s_1 = \sqrt{1 + r^2} \approx 1.1180$, i.e., when the arc center lines 
up with the disk centers as was also the case with the clockwise orbit.
Another simple nonlocalized orbit with period 2 travels along the
diagonal direction (Fig.~\ref{fig-orblg1}b).
It is also born at $s_0=1$ and is allowed for not too large disk
sizes until pruning kills it at $s = \sqrt{2}$ by a flat collision angle.
The other cycles shown in Fig.~\ref{fig-orblg1} are
localized; their symmetry-reduced periods are 1 (Figs.~\ref{fig-orblg1}c
and d) or 2 (Figs.~\ref{fig-orblg1}e and f).

\begin{figure}
\vspace{.5cm}
\centering\leavevmode\epsfysize=6cm\epsfbox{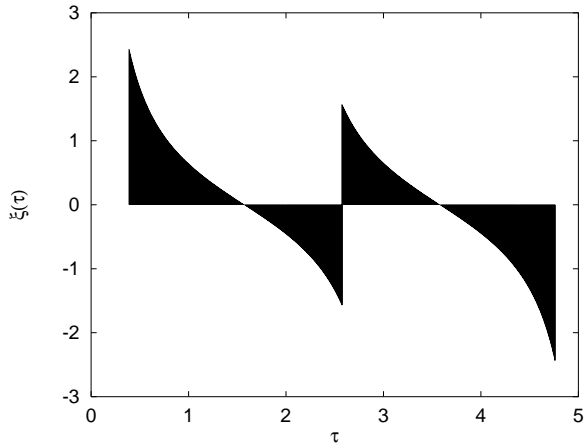}
\caption{The curvature function of the 8-cycle in Fig.~3e for its 
symmetry-reduced period of two collisions at $s=s_1$.}
\label{fig-kpcw2}
\end{figure}

The curvature functions of the period-1 cycles evolve exactly the same
way as described for the clockwise orbit.
The period-2 cycles contain two different collisions, so the
two jumps in $\xi$ are different.
This means that for $s=s_1$ the contributions of the two blocks
to the Lyapunov exponent must cancel each other through a symmetric
arrangement since the integral taken over one block is different from 0
(Fig.~\ref{fig-kpcw2}). 
It is also worth noting that while the period-1 orbits and their islands go
through the evolution described for the clockwise orbit, the 
period-2 cycles show a slightly more complicated behaviour.

For a symmetric period-2 cycle, the $s_1$ values listed in Table I is
specified by the criterion that the cycle is hyperbolic for $s>s_1$,
but this is not the only crossover value between hyperbolic and
elliptic stability.
When increasing $s$ from $s_0$ to $s_1$, the elliptic cycle becomes 
hyperbolic through a period-doubling bifurcation soon after its birth,
but later it restabilizes and remains elliptic up to $s=s_1$ where a
symmetry-breaking equal-period bifurcation (also called Rimmer
bifurcation \cite{Rimmer}) takes place.
This behaviour turns out to be connected to the fact that even-period
cycles has positive stability eigenvalues in weak fields.

So far we have dealt with symmetric orbits built of identical arcs,
although obviously there exist many other cycles with more complicated
symmetries or with no symmetry at all that are stabilized by the field
after their births.
We focused on the symmetric ones not only because they are the
simplest ones to study but also because apparently they are the most
important examples of orbit stabilization in these models in terms
of the size of their ranges of stability and the phase space regions
occupied by their islands. 
In longer cycles with arcs of different lenghts, the focusing effect
of the field in longer arcs in one part of the
orbit may be put off by defocusing in other parts, so the
stabilization is more difficult to achieve and pruning may kill the
orbit before it could stabilize.
In that respect, the symmetry reduces the effective period of the
cycle and makes it more ``sensitive'' to the stabilizing field effects.

\subsection{Multiple collisions}

\begin{figure}
\centering\leavevmode\epsfxsize=12cm\epsfbox{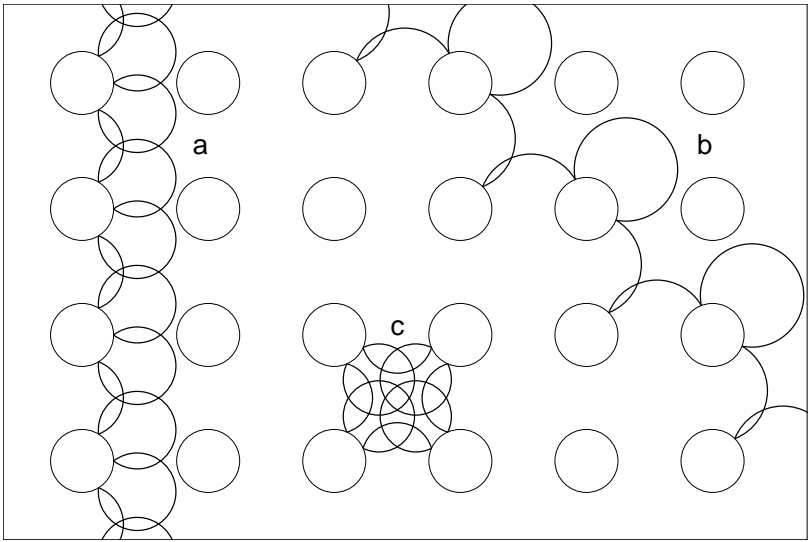}
\caption{Symmetric stable cycles with multiple bounces in the Lorentz
gas; the cyclotron radii are $s=0.615$ (a), 0.82 (b), and 0.565 (c).
These cycles are connected to the regular orbits in Figs.~2a, 2b, and 1,
respectively.}
\label{fig-orblg2}
\end{figure}

In a strong magnetic field, we can obtain new stable cycles from the
symmetric configurations discussed so far by making use of the
the possibility of multiple bounces on the same disk.
The simplest case is when the particle bounces twice on a disk before
it hits another one.
The extra arcs created by the multiple bounces can make
orbit types already extinct in such a strong field reappear by
replacing a simple collision in the original orbit
by one or more arcs with both ends on the same disk. 
Figure~\ref{fig-orblg2} shows three examples in the Lorentz gas. 

\begin{figure}
\centering\leavevmode\epsfysize=6cm\epsfbox{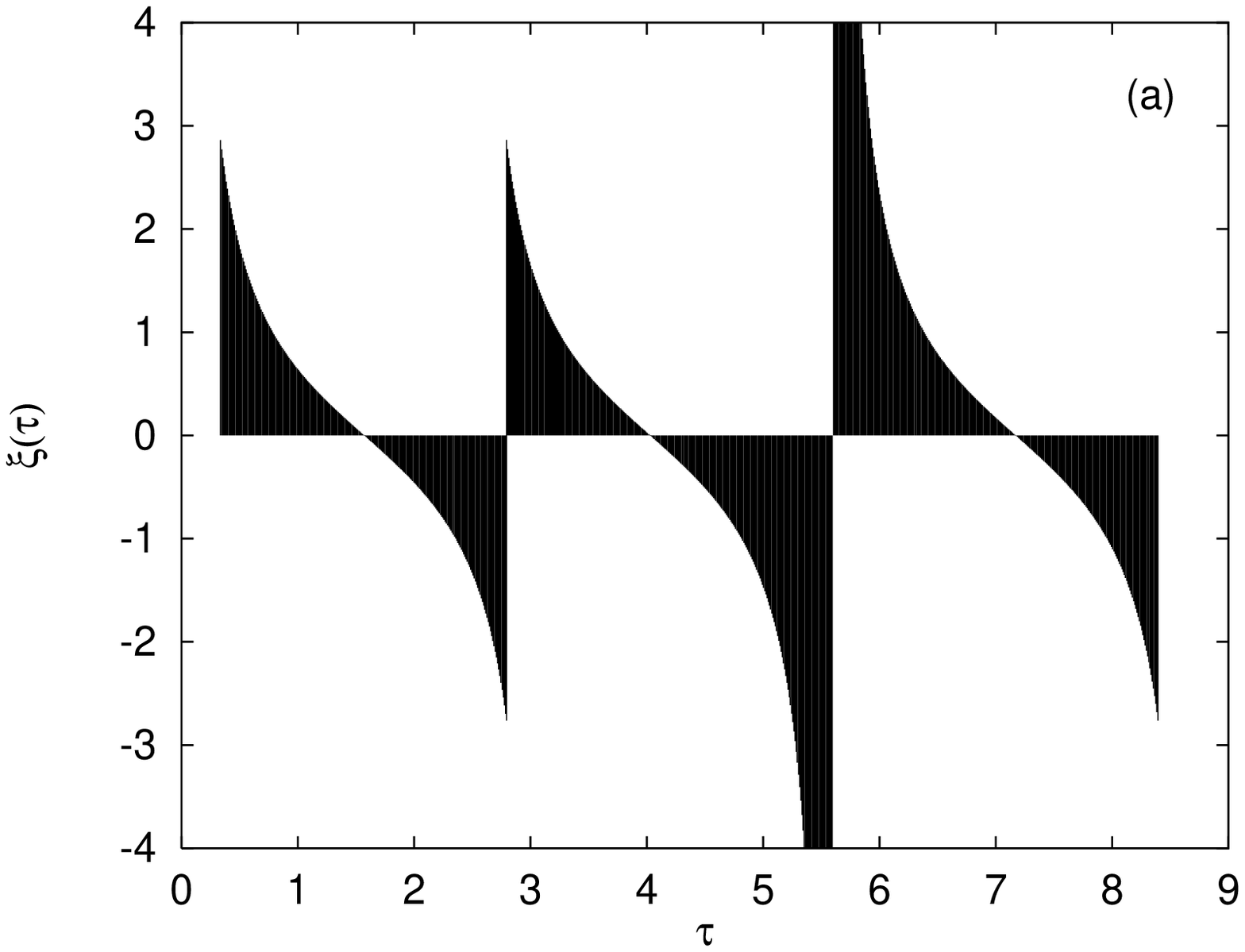}\\[.5cm]
\centering\leavevmode\epsfysize=6cm\epsfbox{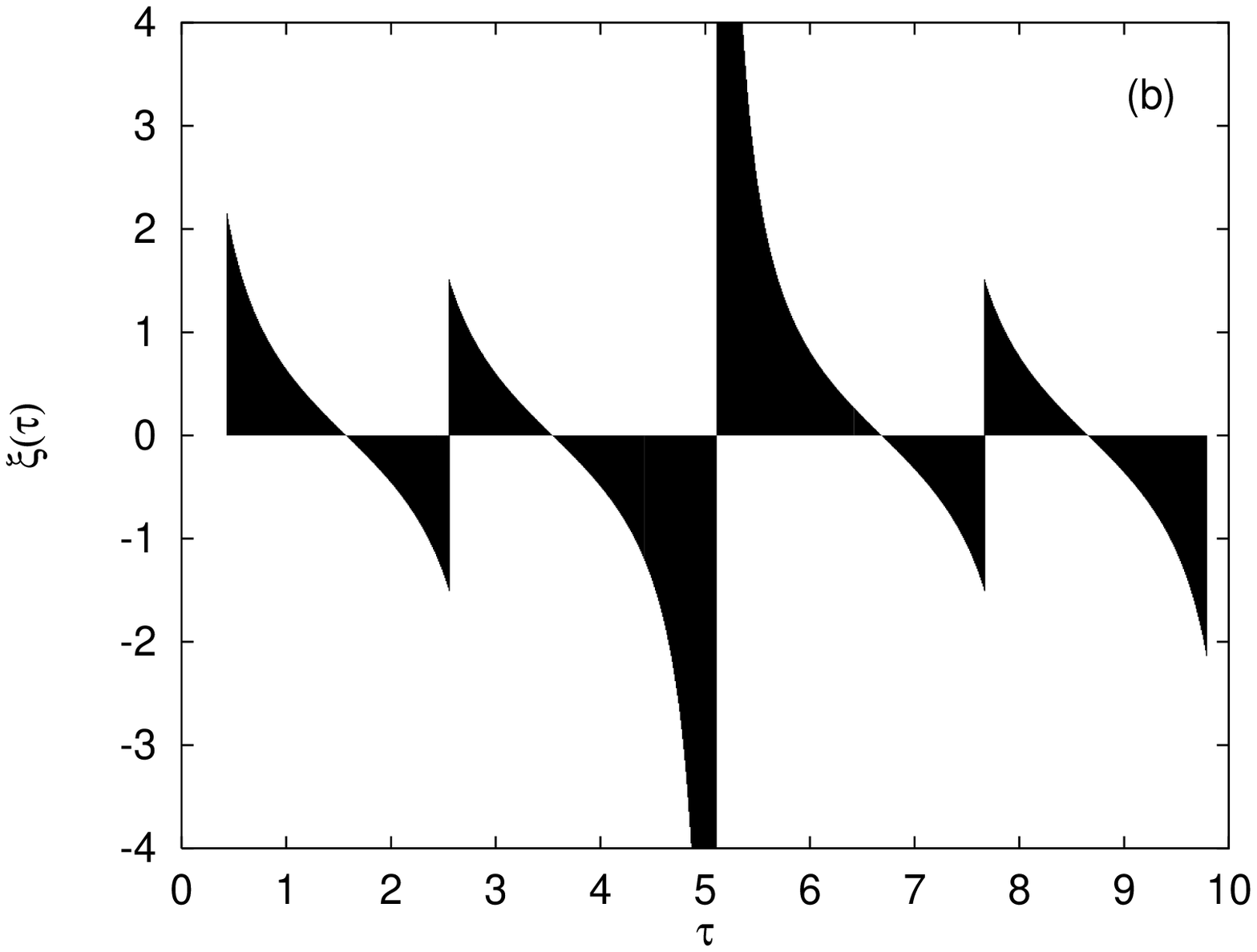}
\caption{The curvature function of the cycles in Fig.~5c (a) and
Fig.~5b (b) when losing stability at their respective field values
$s_1$ (see Table I) for one symmetry-reduced period.
In Fig.~6a, the first block is centered on the node inside as are the
blocks in Fig.~2a for the clockwise orbit at $s=s_1$, while the two 
normal blocks in Fig.~6b could form a curvature function similar to
that in Fig.~4 if the extra middle block were removed.}
\label{fig-kplgx}
\end{figure}

The stabilization mechanism for these orbits is slightly more
complicated than for the ``regular'' orbits because now there are
arcs of different lengths in the cycle.
Looking at the curvature functions of these orbits 
(Fig.~\ref{fig-kplgx}), we can see that at the brink of stabilization,
the block corresponding to the extra arc occupies a
symmetric position with respect to the singularity inside so that
its contribution to the Lyapunov exponent is 0 and the values
$\xi_j$ and $\xi_{j+1}$ at its ends are equal. 
This means that by removing the extra block we could still stick the
other blocks together as if they belonged to a regular orbit with
a similar symmetry: e.g., the period-3 orbit in
Fig.~\ref{fig-orblg2}b, which is a version of the period-2 travelling
orbit of Fig.~\ref{fig-orblg1}b, has two normal blocks that could form
in themselves a curvature function similar to regular period-2 orbits
(cf.\ Fig.~\ref{fig-kpcw2}).

The curvature function also allows us to calculate easily the
stability of special orbits consisting of
collisions with one disk only, rotating around it (usually
quasiperiodically) forever. 
In these orbits, each arc and each collision is equivalent so they can
be treated as period-1 cycles. 
As we have seen, the centered position of the blocks in the curvature 
function automatically satisfies the periodicity condition in this case 
due to the special relationship between the arc length and the impact angle.
This implies that the Lyapunov exponents of
these orbits are always 0, i.e., they are marginally stable at any
field strength where they can exist. 
This ``insensitivity'' can be connected to the fact that the 
islands of such orbits have {\em smooth} boundaries: they lack the
complicated structure seen around the usual KAM-islands.

\section{Stable islands and transport properties}
\label{islands}

\subsection{The effects of hard-wall pruning}

In the Poincar\'e map, an elliptic cycle is surrounded by an island of
bounded motion.
The orbits inside the island visit the disks in the same sequence as
the central cycle remaining inside a certain domain, both in
configuration space and phase space, that they cannot escape.
The island is born at $s=s_0$ together with the corresponding orbit,
and at first it grows in size with decreasing field strength.
It is still there at $s=s_1$ when the cycle in its center becomes
hyperbolic, but this event starts a bifurcation sequence that leads to
the breakup and gradual dissolution of the island in the chaotic see.
The island disappears completely at a cyclotron radius $s_2$ slightly
larger than $s_1$. 

In a certain way, this scenario is similar to those observed in smooth
Hamiltonian systems; however, the hard-wall property, in the form of
pruning, plays a key role in the evolution of the islands. 
This means that the size of an island at a specific field value and the 
details of its disappearance when changing the field may not be solely
the results of the 
``natural'' smooth evolution of its orbits as in smooth systems:
sudden changes in the disk sequences of neighbouring orbits  
limit the room to live for an island and its neighbourhood.
If the outermost orbits of the island should cut through the
interior of other disks in following the disk sequence characterising
the island, then they are removed from the system, and a
well-behaving inner orbit becomes the actual border of the island.
This pushes the edges of the chaotic see closer to the central orbit
and can also limit the width of the border region around
the island where the hierarchy of small island chains and cantori 
can exist.

The influence of hard-wall pruning can clearly be seen in the example
of the orbit in Fig.~\ref{fig-orblg1}d, which is a ``enlarged'' version
of the orbit of Fig.~\ref{fig-orblg1}c connecting next-nearest neighbour
disks instead of nearest neighbours. 
The enlarged cycle has to avoid ``unnecessary'' disks so it is much
more sensitive to pruning.
In fact, it is killed sooner by pruning, when $s$ is increased from
$s_0$, than it could reach destabilization at $s=s_1$. 
This means that the whole island disappears well before the smooth
development of its orbits could make that happen.
In special cases, the onset of hard-wall pruning can even coincide with
the birth of a cycle so that the island cannot develop at all.
For example, the enlarged version of the travelling period-1 orbit 
proceeding along the diagonal line of the unit cells
is born as a pair of two cycles at $s=\sqrt{2}d$ with arcs just 
touching another disk off that diagonal.
The elliptic member of the pair, with arcs shortening when $s$ increasing, 
should cut through that disk which means it is forbidden together with
its island; meanwhile, the hyperbolic cycle moves away from the
disturbing disk so it exists for a certain field range.

\subsection{Trappings near the islands}

The universal orbit structure of smaller islands and cantori around
the edge of an island can affect the ``free'' orbits of the chaotic sea.
Although trajectories started outside the area of an island cannot step 
into it, they can reach its immediate vicinity, through the 
holes of the cantori, where they get trapped for a while until they find
their way out again. 
This stickiness of the island leads to qualitative changes in the
dynamical properties, like the decay of correlations, that
are typical in nonhyperbolic systems. 
The nonhyperbolic effects should be considered when applying the model 
to explain various transport phenomena.
Its consequences from the point of view of chaotic scattering
(together with the evolution of the island for the analogous triangular
orbit) in the three-disk model were discussed in Ref.~\cite{3disk-magn}. 

In the Lorentz gas, the stable orbits can affect the diffusion properties
by trapping other ``free'' trajectories in their vicinity for a while,
thus raising the possibility of anomalous diffusion \cite{KZ-lgmg}.
If a trajectory enters the neighbourhood of the island around a
travelling orbit, then the trapping means moving along with the orbits of
the island at a constant average speed.
On the other hand, in a trapping near a localized orbit the wandering
particles become temporalily localized too.
(For certain ranges of the field strength, the two types of trappings can
even coexist as is the case, e.g., for the orbits of
Figs.~\ref{fig-orblg1}e and \ref{fig-orblg2}b: it can be seen from Table
I that around $s \approx 0.83$ both cycles are stable.)
Whether these trappings can actually modify the character of the
diffusion by forcing a super- or sublinear time dependence on 
the mean square displacement depends on the value of the exponent in
the power law describing the distribution of trapping times.
This question can be approached by using the formalism of L\'evy walks as
in Ref.~\cite{Levy-walks}.  
The results indicate that a typical exponent $\gamma$ in
the trapping time distributions associated with stable islands
($1 < \gamma < 2$, see Ref.~\cite{KAM}) 
leads to superlinear diffusion in the case of travelling
orbits but the localized islands cannot alter the linear
behaviour (although they can decrease considerably the value of the
diffusion constant).
This also implies that the diffusion is superlinear for the coexistence
of the two trapping types too.

All this shows that from a theoretical point of view, it is important to 
know what kind of stable orbits (if any) are present in the system at a
given value of the magnetic field.
This does not necessarily mean in a particular application of the
model, especially in connection with experiments, that the
presence of an island is actually felt in the transport properties,
since the ratio of trajectories affected by the island on the
characteristic time scale of the experiment can be negligible if the
island does not occupy a large area in phase space.
It is worth noting that the islands associated with some of 
the cycles of Figs.~\ref{fig-orblg1} and \ref{fig-orblg2}  
in the Lorentz gas are very small indeed, 
therefore they are easy to miss on a usual phase
space portrait unless we know where to look for them. 
The detailed knowledge of the stabilization mechanism, based on
the properties of the curvature function, can help us in finding them.

\section{Conclusions}
\label{sec-concl}

In the present paper we extended the horocycle description of orbit
separation to the case of billiards in magnetic field. 
We demonstrated that this formalism provides a convenient alternative
for the usual tangent map calulations as a tool to investigate
stability properties of orbits.
In particular, the form of the curvature function suggests which type
of orbits can be stabilized by the field, and this gives an edge to
our method in searching for stable islands compared to plotting a
simple phase space
portrait where the small-size islands can easily be overlooked.
As a simple result, we obtained that all the symmetric periodic orbits
built of arcs of the same length are born stable at a field
strength where the cyclotron diameter takes on the desired value.
We also established the connection between the eigenvalues of elliptic
orbits and the integral of the curvature function.

We identified the most prominent examples in the four-disk model
and in the Lorentz gas and studied their evolutions, also discussing
the consequences of the hard-wall property.
The details concerning which orbits are stabilized by the magnetic
field may
depend on the disk size $r/d$ in the billiard because of the interplay
between smooth orbit evolution and hard-wall pruning.
The longer, more complicated orbits or the enlarged versions of the
simple symmetric orbits may be allowed for a small disk size to go
through their ``natural'' smooth evolution while they may not even be 
present for larger disk sizes because of the pruning.
For weaker field values, when $s$ is considerably larger than $d$,
an elliptic orbit should avoid many other disks in order to form the long arcs
necessary for stabilization, which makes their presence rather unlikely;
therefore, one expects that the creation or 
disappearance of most orbits in weak fields take place without
stabilization.

  From this point of view, it is an interesting open question if 
there is a critical value $s_c$ of the cyclotron radius, depending on
$r$, so that for $s > s_c$ all the existing orbits are unstable in the
system.
For chaotic scattering in the three-disk model, such a
critical field apparently exists \cite{3disk-magn}, so we expect a
similar property in the four-disk model too, but in the Lorentz gas
one cannot rule out the possibility of some unlikely cycles built of 
long arcs in weak field extending over many unit cells and still
avoiding disrupting collisions with ``wrong'' disks. 
(This question is also connected to the existence of collision-free
cycles of closed circles in weak fields.)

The idea of the curvature function and its use as a tool to study 
stability properties can be extended to models quite 
different from the ones discussed here. 
Thermostatted billiard models in external {\em electric} field have drawn
large interest recently from the point of view of the understanding of 
irreversibility and transport in nonequilibrium systems \cite{lg-elec};
in such systems one can use 
the equations of motion for the velocity direction to
derive the time evolution of $\kp$ explicitely. 
As for smooth potentials, the formulae
for the time evolution of the monodromy matrix of Eckhardt and Wintgen 
\cite{monodrom} can be modified to accomodate the effects of an external
magnetic field in the quantity $K$ appearing in the Jacobi equation.
The latter approach is also suitable to treat models, both smooth and 
hard wall, with inhomogeneous fields.

Finally, it is worth noting that the curvature $\kp$ plays a very
important theoretical role in semiclassical calculations using
periodic orbits and the Gutzwiller trace formula. As was shown in
Ref.~\cite{vgdet}, the convergence of the method can be 
drastically improved by
extending the dynamics by adding the tangent space to it so that the
resulting evolution operator becomes multiplicative along the orbits.
The addition of the tangent space dynamics means following the
evolution of $\kp$ as a passive function depending on the ``real''
dynamical variables.
Then taking the trace of the evolution operator requires not only
looking at periodic orbits but also prescribing the periodicity
of $\kp$ for the orbit, which is just the condition we used in our 
study of stability properties.

{\em Added note.} While finishing the manuscript, the author became
aware of Ref.~\cite{tas} dealing with the behaviour of nearby
trajectories in general magnetic billiards on Riemannian surfaces and 
presenting similar results, obtained independently
of the present paper, concerning the evolution of $\kp$.
In particular, our Eq.~(\ref{eq-kp1-free}) is a solution of Eq.~(50) 
of Ref.~\cite{tas} reduced to the special case of Euclidean billiards
in a homogeneous magnetic field, while Eq.~(49) there, describing the
jump of $\kp$ at collisions, agrees with our Eq.~(\ref{eq-kp1-jump}).

\acknowledgments

The author is indebted to G. Vattay for a series of illuminating 
discussions, especially for calling the author's attention to the
Bunimovich---Sinai curvature and pointing out the relevant references.
Interesting discussions with L. Bunimovich, H. W. Capel, J. S. W. Lamb,
and L. Rondoni are also acknowledged.
This work has been done as part of a research project financed by the 
Foundation for Fundamental Research on Matter in The Netherlands 
(project number: 93BR1099); it has also been partially supported by
the Hungarian Scientific Research Foundation (grant numbers: 
OTKA F4286, F17166 and T17493).


\begin{references}
\bibitem{rem-coin}
In our case, the disappearance of the hyperbolic clockwise 
orbit coincides with the birth of the counterclockwise orbit, so in
some sense, these cycles are the two sides of the same coin in this system.
\end{references}
\end{document}